\documentclass{iau307}
\usepackage{graphicx}
\usepackage{natbib}
\usepackage{url}
\usepackage{dtklogos}
\bibpunct{(}{)}{;}{a}{}{,}

\title[An attempt of seismic modelling of $\beta$ Cephei stars in NGC\,6910] 
{An attempt of seismic modelling of $\beta$ Cephei stars in NGC\,6910}

\author[D. Mo\'zdzierski, Z. Ko{\l}aczkowski and E. Zahajkiewicz]   
{D. Mo\'zdzierski, Z. Ko{\l}aczkowski
 \and E. Zahajkiewicz}

\affiliation{Astronomical Institute, University of Wroc{\l}aw, Kopernika 11, 51-622 Wroc{\l}aw \\ email:
{\tt mozdzierski@astro.uni.wroc.pl}}

\pubyear{2014}
\volume{307} 
\pagerange{}
\jname{New windows on massive stars: asteroseismology, interferometry, and spectropolarimetry}
\editors{G. Meynet, C. Georgy, J.H. Groh \& Ph. Stee, eds.}

\begin{document}

\maketitle

\begin{abstract}We present preliminary results of seismic modelling of $\beta$ Cephei-type stars in NGC 6910 based on simultaneous
photometric and spectroscopic observations carried out in 2013 in Bia{\l}k\'ow (photometry) and Apache Point
(spectroscopy) observatories.

\end{abstract}

\firstsection 
\section{Introduction}
The up-to-date seismic modelling of selected bright $\beta$ Cephei stars (see, e.g. \citealt{AertsThoul03},
\citealt{Pamyatnykh04}, \citealt{Dupret04}, \citealt{Daszynska10}) already
brought some constraints on the convective overshooting from the core and an indication that cores in massive stars rotate faster than
their envelopes. The perspectives of the asteroseismology of these stars are therefore promising.

A new way of using asteroseismology is
modelling many stars of the same pulsation type simultaneously which is called ensemble asteroseismology. This type of
asteroseismology can be done e.g.~for members of an open cluster. A prerequisite of a successful asteroseismology is mode
identification and determination of some global stellar parameters which can be done much easier for stars in open clusters than for
field objects. This is because cluster membership makes that some stellar parameters (distance, reddening, age, chemical composition)
can be safely assumed to be the same for all stars, while other parameters (e.g. masses and radii) are strictly related. One of the best
candidates for ensemble asteroseismology is the young open cluster NGC\,6910 in which \citet{Kolaczkowski04}  discovered four $\beta$ Cephei variables.  A new campaign focused on this cluster allowed to detect at least eight $\beta$ Cephei-type members (\citealt{Pigulski08}). 
The frequency spectra of
$\beta$ Cephei stars in this cluster, arranged according to the decreasing brightness (i.e.~mass), show a very interesting progress
of frequencies of excited modes. This is exactly what one could expect for p modes in stars located at the same isochrone in a cluster.
This qualitative result is a strong argument for trying ensemble asteroseismology for this cluster.

\section{Observations and Results\label{SecOne}}

New observations of NGC\,6910 were made in 2013. The photometric observations were obtained in Bia{\l}k\'ow Observatory (Poland)
during 21 nights. These observations were carried out with a 60-cm reflecting telescope and the attached CCD
camera covering 13$^\prime$ $\times$ 12$^\prime$ field of view. About 4000 CCD frames through $B$, $V$, and $I_{\rm C}$
filters of the Johnson-Kron-Cousins photometric system were acquired. Spectroscopic observations were
carried out with the Apache Point Observatory (APO) ARC 3.5-m telescope and the ARC Echelle Spectrograph (ARCES) during five nights.
In total, we have taken 36 spectra of NGC\,6910-18 and single-epoch spectra of two other $\beta$ Cep-type stars:
NGC\,6910-14 and NGC\,6910-16. The spectra have a resolving power of 31500 and cover a range between 3200 and 10000 \AA{}.
Photometric observations were calibrated in a standard way. For each frame, we calculated aperture and profile magnitudes
of the stars using the {\sc Daophot II} package \citep{Stetson87}, and then derived differential magnitudes. Spectroscopic
observations were reduced with standard IRAF routines.

We determined $T_{\rm eff}$ and $v\sin i$ of the observed stars (NGC\,6910-14, -16, -18) using our spectra and the BSTAR2006
grid of non-LTE model atmospheres of \citet{Lanz07} and the ROTIN3 program. We used also the UVBYBETA code and
literature $u$, $v$, $b$, $y$, and $\beta$ magnitudes to obtain $\log \mbox{$L/L_{\rm \odot}$}$ and $\log T_{\rm eff}$
of NGC\,6910-14 and NGC\,6910-18. The results are shown in Table 1. In the case of NGC6910-18 we found that the
amplitudes of the two dominating modes 
with frequencies $f_1 =$ 6.1549 d$^{-1}$ and $f_2 =$ 6.3890 d$^{-1}$ remained almost unchanged in comparison to
2005--2007 observations. We performed mode identification for this star with the methods
developed by \citet{Daszynska05} for five stellar models using $B$, $V$, and $I_{\rm C}$ time-series photometry. 
The evolutionary tracks were computed with
the Warsaw-New Jersey evolutionary code adopting the OP opacities, the solar mixture, rotational
velocity $V_{\rm rot} =$ 100 km\,s$^{-1}$, hydrogen abundance $X =$ 0.7, metallicity parameter $Z =$ 0.015 and no overshooting
from the convective core. We identified $f_1$ as an $l =$ 3 mode whereas $f_2$ can be identified as $l =$ 0, 1 or 2.\\

\begin{table}
\begin{center}
\caption{Atmospheric parameters of analysed $\beta$ Cep stars in NGC\,6910.}
\label{tab1}
\begin{tabular}{c|cccc|}\hline 
\textbf{Star} & \textbf{$\log T_{\rm eff}$ $^1$} & \textbf{$\log T_{\rm eff}$ $^2$} & \textbf{$\log \mbox{$L/L_{\rm \odot}$}$} & $v\sin i$ [km\,s$^{-1}$]\\ 
\hline
NGC\,6910-14 & 4.447 & 4.443 & 4.182 & 125\\
NGC\,6910-16 & 4.447 & --- & --- & 149\\
NGC\,6910-18 & 4.398 & 4.400 & 4.025 & 94 \\
\hline
\end{tabular}
\end{center}
\vspace{1mm}
\scriptsize{
{\it Notes:}\\
$^1$Based on our spectroscopy. \\
$^2$Based on Str\"omgren photometry.}
\end{table}

This work was supported by the NCN grant No. 2012/05/N/ST9/03898 and has received funding from the EC
Seventh Framework Programme (FP7/2007-2013) under grant agreement no. 269194. The work is based on observations obtained with
the Apache Point Observatory 3.5-meter telescope, which is owned and operated by the Astrophysical Research Consortium.
Some calculations have been carried out in Wroc{\l}aw Centre for Networking and Supercomputing
(http://www.wcss.wroc.pl), grant No. 219.  We thank Dr.\,J.\,Jackiewicz for making available telescope time on
the APO ARC 3.5-m telescope and for his support during observations. We thank A.\,Pigulski, P.\,Bru\'s, G.\,Kopacki and 
P.\,\'Sr\'odka for making some observations of NGC\,6910 in Bia{\l}k\'ow.

\bibliographystyle{iau307}
\bibliography{MyBiblio}

\end{document}